\documentclass[print,showpacs,oneside,twocolumn,superscriptaddress,prl,amsmath,amssymb]{revtex4}
\usepackage{cases}
\usepackage{amsmath}
\usepackage{amssymb}
\usepackage{amsfonts}
\usepackage{amssymb}
\usepackage{dcolumn}
\usepackage{bm}
\usepackage{hyperref}
\usepackage{graphicx}
\usepackage{epstopdf}

\begin{document}
\title{Landau-Zener-Stuckelberg interferometry of a single electron spin in a noisy environment}
\author{Pu Huang}
\author{Jingwei Zhou}
\author{Fang Fang}
\author{Xi Kong}
\author{Xiangkun Xu}
%\author{Ya Wang}
%\author{Fazhan Shi}
%\author{Xin Rong}
%\author{D}
\affiliation{Hefei National Laboratory for Physics Sciences at
Microscale and Department of Modern Physics, University of Science
and Technology of China, Hefei, 230026, China}

\author{Chenyong Ju}
\altaffiliation{cyju@ustc.edu.cn}
\affiliation {Hefei National Laboratory for Physics Sciences at Microscale and Department of
Modern Physics, University of Science and Technology of China,
Hefei, 230026, China}

\author{Jiangfeng Du}
\altaffiliation{djf@ustc.edu.cn}
\affiliation {Hefei National Laboratory for Physics Sciences at Microscale and Department of
Modern Physics, University of Science and Technology of China,
Hefei, 230026, China}

\begin{abstract}
We demonstrate quantum coherent control of a single electron spin in a NV center in diamond using the Landau-Zener-Stuckelberg interferometry at room temperature. Interference pattern is observed oscillating as a function of microwave frequency. The decays in the visibility of the interference are well explained by numerical simulation which includes the thermal fluctuations of the nuclear bath which shows that Landau-Zener-Stuckelberg interferometry can be used for probing electron spin decoherence processes.

\end{abstract}


\maketitle
Landau-Zener(LZ) tunneling is a well-known phenomenon associated with
strong-driving. When a two-level system is driven through the avoided level
crossing, LZ tunneling can be controlled to create various superposition of
the energy eigenstates, like a coherent beam splitter. Combining two
consecutive LZ tunneling leads to Landau-Zener-Stuckelberg(LZS) quantum
interference which is analogous to the Mach-Zehnder(MZ) interferometry. LZS
interference has been observed in Rydeberg atoms\cite{1Rydberg
atoms,2Rydberg atoms}, quantum dots contacts\cite{3QPC}, and recently in
mesoscopic superconducting Josephson devices\cite{SCJ,CPB}, ultracold
molecular\cite{ultracold} and Optical Lattices\cite{optic}. LZ tunneling
and LZS interference have also been exploited for quantum states preparation%
\cite{7,8} and manipulation\cite{9,10}. Interaction with the environment
disturbs the coherence of the quantum system and therefore manifests in the
LZS interference pattern\cite{dec1,dec2}. LZS interferometry thus provides
crucial information on the decoherence processes by the environments.

Single electron spins of nitrogen-vacancy centre(NV centre)has been one of
the most popular candidate as a qubit carrier. The high controllability as
well as a favorable coherence time promises room temperature quantum
information processing\cite{coherence1,entangle,read1,coherence2}and high
sensitive magnetometry\cite{mag1,mag2,mag3}. High efficiency initialization
of NV center spin on desired pure state can be done via optical pumping.
Quantum coherent controls of individual spin can be realized using the
conventional microwave pulsed controls. NV center is also an ideal platform
for studying quantum phenomena, especially, the processes in a
strong-driving regime such as anharmonic dynamics\cite{Gigahertz} and the
multifrequency spectra\cite{multi}.

In this letter, we carried out LZS interferometry on a NV centre in high
purity diamond and demonstrated the feasibility of using LZS for quantum
coherent control of single electron spin. In this experiment, we first
realize a coherent beam splitter for electron spin states based on the LZ
tunneling process, this is realized in and then, by repeating such process
twice under different microwave frequency, quantum interference known as
Stuckelberg oscillation is observed. The decays in the interference fringes
agrees well with numerical simulations taking into account the hyperfine
coupling of the electron spin with the surrounding nuclear spin bath. Our
study shows that the thermal fluctuations of the nuclear spins is the
dominate cause of observed coherence lose at room temperature.

\begin{figure}[tbp]
\begin{center}
\includegraphics[width= 1\columnwidth]{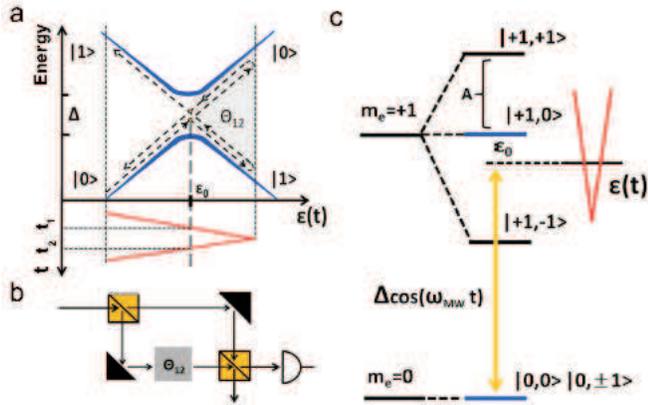}
\end{center}
\caption{(color online). Scheme for realizing an LZS interferometer on a
single spin of the NV center in diamond (a) Energy diagram of a two-level
LZS interferometer: start at $|0\rangle $, the system is driven through the
avoided crossing(at $t_{1}$) and split into a superposition of $|0\rangle $
and $|1\rangle $ via LZ tunneling; after accumulating a relative phase $%
\protect\theta _{12}$, $|0\rangle $ and $|1\rangle $ interfere at the
avoided crossing(at $t_{2}$). (b) An MZ interferometer can be used to
describe the LZS interferometer, with LZ tunneling acting as an optical beam
splitter. (c) Realization of the LZ hamiltonian in an NV center: the
two-level system of the LZS interferometer is expanded in electron states $%
m_{e}=0$ and $m_{e}=+1$ with ${}^{14}N$ nuclear spin in the $m_{I}=0$ state
respectively(blue lines), a selective microwave field with frequency $w_{MW}$%
(yellow double arrow) with strength $\Delta $ can act as the minimum energy
separation after going into the rotating frame, $\protect\varepsilon _{0}$
in the LZ problem can be tuned directly by the microwave frequency. The
driving field $\protect\varepsilon (t)$ (red folded lines) is applied along
[111] crystal axis. }
\label{fig1}
\end{figure}

LZ tunneling was first studied by Landau and Zener\cite{Ori1,Ori2} and a
description of the variety of phenomenon related to LZ tunneling and LZS
interference can be found in a recent review\cite{26}. The problem rely on a
two-level system, described by the LZ Hamiltonian,
\begin{equation}
H_{LZ}=-\frac{\Delta }{2}\sigma _{x}-\frac{\varepsilon (t)-\varepsilon _{0}}{%
2}\sigma _{z},  \label{eq£º1}
\end{equation}%
which contains a minimum energy separation $\Delta $ (the avoided level
crossing), a time dependent driving field $\varepsilon (t)$ and a offset $%
\varepsilon _{0}$. To implement LZ tunneling, the system is prepared in an
eigenstate of $\sigma _{z}$, denoted by $|0\rangle $, while the driving
field $\varepsilon (t)$ is set to be much larger than $\Delta $. In this way
the initial state $|0\rangle $ is close to one of the eigenstates of $H_{LZ}$%
. Next, $\varepsilon (t)$ is gradually tuned down, and as the system is
swept through the avoided level crossing where $\varepsilon (t)=\varepsilon
_{0}$, $|0\rangle $ undergoes LZ tunneling and is split into a superposition
of $|0\rangle $ and $|1\rangle $. The probability of remaining in $|0\rangle
$ is given by the well-known LZ formula:
\begin{equation}
P_{T}=exp(-\frac{\pi }{2}\delta )  \label{eq£º2}
\end{equation}%
where $\delta =\Delta ^{2}/v$ and $v$ is the sweep velocity which equals to
the value of $d\varepsilon (t)/dt$ at the avoided crossing. The LZS
interferometer can also be conceived as a MZ interferometer,
as shown in Fig. \ref{fig1}(b). Starting with $|0\rangle $, the system undergoes LZ
tunneling at $t_{1}$ where the superposition of the $|0\rangle $ and $%
|1\rangle $ is generated. The system subsequently evolves and accumulates a
relative phase $\theta _{12}$. When $\varepsilon (t)$ is swept back to the
avoided level crossing at $t_{2}$, $|0\rangle $ and $|1\rangle $ interfere. The relative phase $\theta _{12}$ is given by
\begin{equation}
\theta _{12}=\int_{t_{1}}^{t_{2}}E_{01}(t)dt  \label{eq£º3}
\end{equation}%
with $E_{01}(t)$ the energy difference between $|0\rangle $ and $|1\rangle $%
. It is $\theta _{12}$ that gives rise to the interference fringes in the
occupation probability known as Stuckelberg oscillations.

The key issue in LZ tunneling is the realization of an avoided level
crossing in energy states. Nevertheless, due to the large crystal splitting
of the NV center, such a scheme cannot be implemented directly by electron
spin resonance where a formidable strength of microwave field is required.
In our scheme, the avoided level crossing is realized in the rotating frame
where a microwave field can act as the minimum energy separation, similar
method has been used in solid state electron spins \cite{Gigahertz,multi} as
well as in superconducting Josephson devices\cite{Yu}. We consider the NV
center system which includes a spin-one electron spin and a nearby ${N}^{14}$
nuclear spin. The Hamiltonian can be written as
\begin{equation}
H_{NV}=DS_{z}^{2}+g_{e}\mu _{B}B_{z}S_{z}+A_{z}I_{z}S_{z}.  \label{eq£º4}
\end{equation}%
Here $D\approx 2.87$ GHz is the crystal field splitting and $B_{z}$ is the
external magnetic field applied along the $z$ axis ([111] crystal axis) which lifts the degeneracy of electron spin states $|+1\rangle _{e}$ and $|-1\rangle _{e}$. $%
\mu _{B}$ is the Bohr magneton and $g_{e}$ the electron $g$-factor. The third
term of Eq. (\ref{eq£º4}) is the hyperfine coupling between the electron spin and the $%
{}^{14}N$ nuclear spin ($A\approx 2.18$MHz), which contributes an effective
field to the center spin conditioned on the ${}^{14}N$ state (we neglect the
dynamics of ${}^{14}N$ for simplicity.) To realize the corresponding avoided
level crossing in such a system, we first transform to the subspace spanned
by $|0\rangle _{e}$ $|0\rangle _{I}$ and $|+1\rangle _{e}$ $|0\rangle _{I}$
as in Fig. \ref{fig1}(c), which will be denoted by $|0\rangle $ and $|1\rangle $ in
the following. Applying a microwave field $\Delta \cos (w_{MW}t)$
along the $x$ axis selectively excites the transition $|0\rangle $ $%
\leftrightarrow $ $|1\rangle $ when $\Delta \ll A$. In the rotating frame of
$w_{MW}$, $\Delta $ acts as a static field along the $x$ axis. Finally, by
assuming a time dependent field $\varepsilon (t)$ along the $z$ axis with
amplitude smaller than $A_{z}$, the dynamics of $|0\rangle $ and $|1\rangle $
in the rotating frame can be expressed by $H_{LZ}$. By tuning the strength
of the microwave field, one can change the minimum energy separation $\Delta
$, while by tuning the microwave frequency, $\varepsilon _{0}$ can be
controlled.

We describe the experimental set-up for the demonstration of the LZS
interferometry. The experiment is carried out on a home-built confocal
microscope operated at room temperature. The sample is type IIa single
crystal diamond with abundance of nitrogen electron spins less than 5ppb. A
single NV is addressed via a microscope mounted on a piezoscanner by its
fluorescence signals. A Hanbury-Brown-Twiss setup with two photodetectors is
used to ensure the single NV (data not shown). A 532nm laser is used to
initialize and read out the system. To manipulate the electron spin
coherently, a microwave signal is first generated by a ratio signal
generator, then a linear amplifier is employed to enhance the microwave
power output. Finally, a 20$\mu$m diameter copper wire terminated by a $50$%
Ohm resistance is used to radiate the microwave field to the NV center. The
degeneracy between $|+1\rangle_{e}$ and $|-1\rangle_{e}$ is lifted by an
external magnetic field generated by three pairs of Helmholtz coils, with
resolution $\approx 0.01$ Gauss. In the experiment a magnetic field of 5
Gauss is employed. The driving field $\varepsilon(t)$ in $H_{LZ}$ is
generated by an Arbitrary Waveform Generator(AWG) and the signal is directly
sent to the sample via the copper wire. In the experiment, the typical
frequency of the driving field is several kHz which is much smaller than the
microwave frequency, thus only the components along the $z$ axis contribute.
All signals are synchronized by a pulse generator. To build up statistic, we
use typically $10^5$ cycles in a single measurement. $\Delta$ is calculated
using the output power of the amplifier and the amplitude of the driving
field from the output voltage of the AWG. Typically, $\Delta \approx 500$
kHz at 20dBm output while 4V (peak-to-peak value of sine wave) in AWG
corresponds to a driving field amplitude of 1.4MHz.

\begin{figure}[tbp]
\begin{center}
\includegraphics[width= 1\columnwidth]{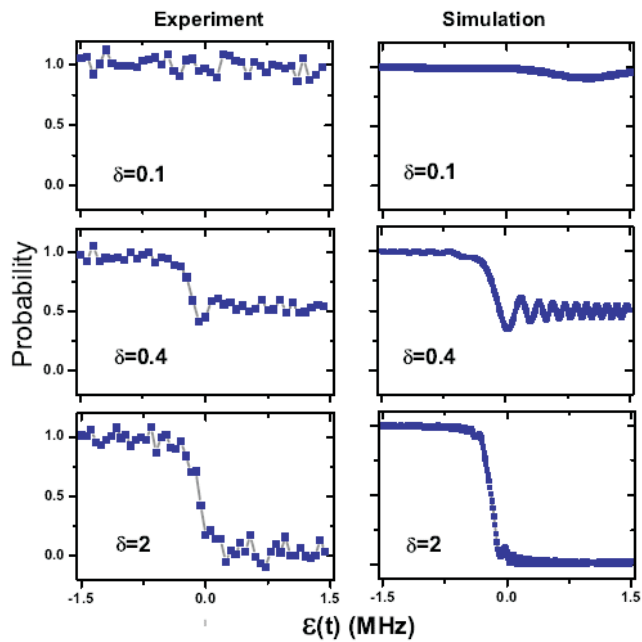}
\end{center}
\caption{(color online). Dynamics of the LZ tunneling. Measured(left) and
simulated(right) dynamics of LZ tunneling under driving field: $\protect%
\varepsilon (t)=\protect\varepsilon $cos($2\protect\pi wt$) for different
values of the adiabaticity parameter $\protect\delta =\Delta ^{2}/\protect%
\varepsilon w$. To measure this curve, we set $\protect\varepsilon =$1.5MHz
and $\Delta =$0.11KHz and used different $w$. For the case $\protect\delta %
\ll 1$ , the system remains in $|0\rangle $ while for $\protect\delta \gg 1$%
, the system evolves adiabatically with the driving field. }
\label{fig2}
\end{figure}
\begin{figure}[tbp]
\begin{center}
\includegraphics[width= 1\columnwidth]{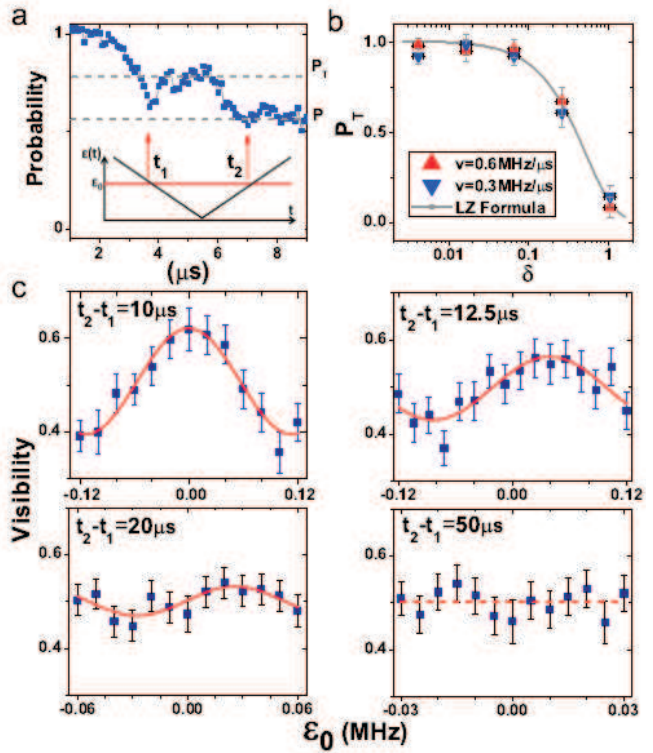}
\end{center}
\caption{(color online). LZS interferometer. (a) Dynamics of the LZS
interference process driven by $\protect\varepsilon(t)=vt$ (inset). gray
lines indicate the probability of being in the state $|0\rangle$ after
passing through the first ($P_{T}$) and the second avoided level crossing ($P
$). (b) $P_{T}$ as a function of the adiabaticity parameter $\protect\delta$
(triangles) for different sweeping velocities. The result is in good
agreement with the LZ formula (2). (c) Measured Stuckelberg oscillation
curves (blue squares) for different evolution durations $t_{2}-t_{1}$. The
red line represents a fit to a cosine in order to obtain the visibility. The
dashed line in last sub-diagram is a guide for the eyes since the visibility
is nearly zero. In measuring this curve, we have adjusted the microwave
power to bring $P_{T}$ close to $1/2$, so that the oscillation amplitude is
a maximum according to the classical LZ theory. The measured visibility is
defined as $P/4P_{T}(1-P_{T})$. }
\label{fig3}
\end{figure}

Next, we describe the demonstration of the LZS interferometer. As
preparation, we first accomplished LZ tunneling that can act as a beam
splitter for electron spin states. Fig. \ref{fig2} shows the measured and simulated
LZ tunneling dynamics under a driving field $\varepsilon $cos$(2\pi wt)$ for
different values of the adiabaticity parameter $\delta =\Delta
^{2}/\varepsilon w$. The simulation is based on $H_{LZ}$ and the environment
is not taken into account. By tunning the adiabaticity parameter, the
tunneling probability can be controlled. Such a beam splitter can be further
exploited to construct the LZS interferometer. According to the classical
LZS interference theory\cite{26}, beginning in $|0\rangle $, the probability
$P$ of the system keeping in the same state after undergoing two consecutive
LZ tunneling is given by
\begin{equation}
P=4P_{T}(1-P_{T})sin(\Phi )^{2}.  \label{eq£º5}
\end{equation}%
%You haven't defined P! It is only defined in the figure.
Here $P_{T}$ is the LZ tunneling rate, and $\Phi =\Phi _{S}+\theta _{12}$
with $\Phi _{S}$ coming from Stokes phase. Fig. \ref{fig3}(a) shows the
characteristic dynamics of the LZS interference process, while $P_{T}$ as a
function of sweeping velocity $v$ is plotted in Fig. \ref{fig3}(b). To extract out
the quantum coherence term in (4), we defined the interference visibility $V$
as $V=P/4P_{T}(1-P_{T})$. Without noises, $V$ is expected oscillating
between 0 and 1 as a function of $\theta _{12}$. However, as shown by the
observed Stuckelberg oscillation(Fig. \ref{fig3}(b)), the visibility of interference
fringes decrease as increasing the duration of the interference process.
Such an effect clearly indicates a loss of quantum coherence between $%
|0\rangle $ and $|1\rangle $.

\begin{figure}[tbp]
\begin{center}
\includegraphics[width= 1\columnwidth]{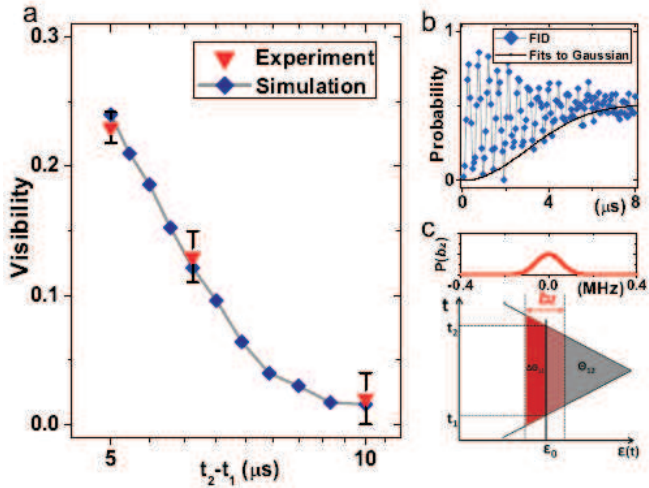}
\end{center}
\caption{(color online). Decoherence in the LZS process. (a) Measured (red
triangles) and simulated (blue rectangle) visibility as a function of
duration between the two LZ tunneling events. Agreement between theory and
the experiment confirms that it is the thermal fluctuations of the nuclear
spin bath that cause the decay of the visibility. (b) Measured FID signal,
data is fitted to Gaussian decay(black line). (c) An intuitive picture can
be employed to understand these observations. The fluctuations of nuclear
spins produce an effective magnetic field $b_z$. It changes $%
\protect\theta_{12}$ which is proportional to the area of the gray regime,
with the result that the phase information is washed out. }
\label{fig4}
\end{figure}

To understand the observed decay in the interference pattern, the
environment must be taken into account. In high purity single crystals,
where the abundance of nitrogen electron spins is less than 5ppb, the main
source of decoherence comes from the dipolar interaction with the $^{13}C$
nuclear spins \cite{decoherence}. This interaction, together with the
dynamics of the nuclear spin bath can be expressed by $\hat{b}_{z} |1\rangle \langle 1| + H_{bath}$
%$\hat{b}_{z}\sigma_{z}/2 + \hat{b}_{z}/2 + H_{bath}$
in the subspace of the LZS
interferometer. Here $\hat{b}_{z}$ is the coupling to the nuclear spin bath,
which can be written as $\Sigma _{j}\mathbf{A}_{j} \cdot \mathbf{I}_{j}$,
with $\mathbf{A}_{j}$ the coupling coefficient for the $j$th nuclear spin $%
\mathbf{I}_{j}$ (the coupling strength is of the order of kHz). The
fluctuation perpendicular is negligible since it is too weak to cause the
spin-flip relaxation. $H_{bath}$ contains the dynamics of the bath, which
includes the Zeeman splitting (several kHz) of the nuclear spins in the
external magnetic field and the dipolar interaction between nuclear spins
(of the order of Hz). During the interference process, which occurs within
tens of microsecond, the dynamics of the bath are negligible. We therefore
expect that only the statistical fluctuations arising from the random
orientations of the $^{13}C$ nuclear spins at room temperature contribute to
the interference process. These fluctuations follow a Gaussian distribution $%
exp(-b_z^{2}/2\beta^{2}) $\cite{NVdec1}, where $\beta$ can be directly
extracted from the FID measurement (Fig. \ref{fig4}(b)). It is found that $\beta =
0.056$kHz for the NV center under study.

Based on these considerations, numerical simulations were performed, with
the measured and simulated results shown in Fig. \ref{fig4}(a). Good agreement
between experiment and theory clearly establishes the decay mechanism as
being due to nuclear spins. One also can capture the essence of the
observations through the intuitive picture presented in Fig. \ref{fig4}(c). The phase
giving rise to the interference fringes comes from the energy accumulated
between two LZ tunneling points and is proportional to the duration of
interference process multiplied by the amplitude of the driving field. The
presence of the effective field $b$ can change the position of the avoided
level crossing and therefore cause fluctuations in $\theta_{12}$. As a
result, the oscillations are washed out and the visibility decreases. This
effect becomes more serious as the duration of the total process increases.

In conclusion, we have demonstrated a beam splitter of spin states of the NV
centre at room temperature based on LZ tunneling, Our results showed that
the tunneling probability is only given by the adiabaticity parameter $\delta
$ which agrees with the original prediction of LZ formula. Combining two
such beam splitters, LZS interferometer is realized and the Stuckelberg
oscillation is observed, the decays in visibility of the interference
fringes at room temperature is caused by thermal fluctuation of nuclear
spins and agrees well with numerical simulations. Our work establishes the
feasibility of using LZS interferometry for quantum coherent control and for
probing decoherence processes of single spin in NV center.

We thank W. Yao and D. Culcer for helpful discussion and reading the paper.
This work was supported by the National Natural Science Foundation of China
(Grant No. 91021005), the CAS, and the National
Fundamental Research Program (Grant No. 2007CB925200).


\begin{references}
\bibitem{1Rydberg atoms}S. Yoakum {\it et al.}, Phys. Rev. Lett. \textbf{69}, 1919 (1992).
\bibitem{2Rydberg atoms}Baruch,M.C and T.F.Gallagher, Phys. Rev. Lett. \textbf{68}, 3515 (1992).
\bibitem{3QPC}Gorelik L.Y. {\it et al.}, Phys. Rev. Lett. \textbf{81}, 2538 (1998).
\bibitem{SCJ} W. D. Oliver {\it et al}, Science \textbf{310}, 1653 (2005).
\bibitem{CPB} S. Sillanpaa {\it et al.}, Phys. Rev. Lett. \textbf{96}, 187002 (2006).
\bibitem{ultracold} M. Mark {\it et al.}, Phys. Rev. Lett. \textbf{99}, 113201 (2007).
\bibitem{optic} S. Kling {\it et al.}, Phys. Rev. Lett. \textbf{105}, 215301 (2010)
\bibitem{7}D. J. Reilly {\it et al.}, Science \textbf{321}, 817 (2008).
\bibitem{8} H. Ribeiro {\it et al.}, Phys. Rev. Lett. \textbf{102}, 216802 (2009).
\bibitem{9} J. R. Petta {\it et al.}, Science \textbf{327}, 669 (2010).
\bibitem{10} Guozhu Sun {\it et al.}, Nature communications \textbf{1(5)}, 51 (2010).
\bibitem{dec1} D. M. Berns {\it et al.}, Phys. Rev. Lett. \textbf{97}, 150502 (2006).
\bibitem{dec2} M. S. Rudner {\it et al.}, Phys. Rev. Lett. \textbf{101}, 190502 (2008).
\bibitem{coherence1}M. V. Gurudev Dutt {\it et al.} Science 316, 1312 (2007).
\bibitem{entangle}P. Neumann {\it et al.}, Science 320, 1326 (2008).
\bibitem{read1} P. Neumann {\it et al.}, Science 329, 542 (2010).
\bibitem{coherence2} G. Balasubramanian et al., Nat. Mater. 8, 383 (2009).
\bibitem{mag1} J. R. Maze {\it et al.}, Nature 455, 644 (2008).
\bibitem{mag2} G. Balasubramanian {\it et al.}, Nature (London) 455, 648 (2008).
\bibitem{mag3} G. de Lange {\it et al.}, Phys. Rev. Lett. \textbf{106}, 080802(2011).
\bibitem{Gigahertz} G. D. Fuchs {\it et al.}, Science \textbf{326}, 1520(2009).
\bibitem{multi} L. Childress {\it et al.}, Phys.Rev.A \textbf{82}, 033839(2010).
\bibitem{Ori1} Landau,L.D., Phys.Z \textbf{2}, 46(1932).
\bibitem{Ori2} Zener,C, Proc.R.Soc.London A \textbf{137}, 696(1932).
\bibitem{26} S. Shevchenko, S. Ashhab, and F. Nori, Phys. Rep. 492, 1 (2010).
\bibitem{Yu} G. Sun {\it et al.}, arXiv:1010.5897 .
\bibitem{decoherence} L. Childress {\it et al.}, Science \textbf{314}, 281(2006).
\bibitem{NVdec1} R. Hanson {\it et al.}, Science \textbf{320}, 352(2008).



\end{references}
\end{document}